# A standing Leidenfrost drop with Sufi-whirling

*Jinlong Yang[1], Yong Li[2], Yue Fan[3], Longquan Chen[4], Dehui Wang[1], Xu Deng[1]\**

1. Institute of Fundamental and Frontier Sciences, University of Electronic Science and Technology of China, Chengdu, China
2. Digital Media Art Key Laboratory of Sichuan Province, Sichuan Conservatory of Music, Chengdu, China
3. School of Materials Science and Engineering, Sun Yat-sen University, Guangzhou, China.
4. School of Physics, University of Electronic Science and Technology of China, Chengdu, China

Corresponding author: dengxu@uestc.edu.cn (X.D.)

**Abstract**

The mobility of Leidenfrost drop has been exploited for the manipulation of drop motions. In the classical model, the Leidenfrost drop was levitated by a vapor cushion, in the absence of touch to the surface. Here we report a standing Leidenfrost state on a heated hydrophobic surface where drop stands on the surface with partial adhesion and further self-rotates like Sufi-whirling. To elucidate this new phenomenon, we imaged the evolution of the partial adhesion, the inner circulation, and the ellipsoidal rotation of the drop. The stable partial adhesion is accompanied by thermal and mechanical equilibrium, and further drives the development of the drop rotation.

When deposited on a hot surface, free water droplets either undergo explosive boiling or keep in a stable mild state, hovering over a thin air cushion. The latter phenomenon was first observed and investigated by scientist Johann Gottlob Leidenfrost, who observed a relatively long-lasting droplet on a hot pan.[1] The thin air layer enables the drop with high mobility; to date, various Leidenfrost phenomena have been observed and investigated, including oscillation, rolling, trampolining, and jumping up, among others.[2-7] The fruitful outcome originates from the highly mobile yet constrained droplet, stimulated by both thermal and mechanical vibrations.[8] The air cushion-induced vibration promotes the oscillation of the drop, the form of which is constrained by the drop size.[2,3] The shear force of the airflow in the cushion stimulates a size-dependent vertical circulation inside the drop and promotes drop rolling.[4] The pressure vibration of the air cushion promotes trampolining of the drop.[5] At the life end of the Leidenfrost drop, the impurity induced vertical jump of the drop, triggered by the drastic interaction between the droplet and the surface.[6]

In contrast to conventional Leidenfrost dynamics, we observe a new phenomenon that has not been studied. A drop on a heated hydrophobic surface exists in a stable state with partial adhesion to the surface, namely, a standing Leidenfrost drop. In this state, the drop further evolves to horizontal deformation and rotates like Sufi-whirling. We unravel the physical reasons behind the stable partial adhesion and self-rotation mechanism by analyzing the thermal and mechanical equilibrium, the evolution of the drop rotation, the origination of the circulation drive, and the constraints for this phenomenon.

When a droplet of 20 μL was placed gently on a heated smooth surface, the outcome was dramatically influenced property of the substrate and the temperature.[9,10] In the classical boiling state, nucleating boiling, transitional boiling, and film boiling (namely the Leidenfrost state) take place sequentially with the elevation of temperature (Figure S1, Supplementary note 1). When the substrate was replaced by an overhanging hydrophobic slide (setup in Figure S2), we observed an additional state, that is, a standing Leidenfrost drop with a 'foot' (partial adhesion) on the surface. Figure 1a shows the evolution of a drop placed on the hot hydrophobic surface at different temperatures. With the elevation of the temperature, the water drop starts to boil in a drastic behavior, which was termed transitional boiling. Further increasing the temperature sees that the droplet undergoes a short boiling with vibration and then quiets down on the surface, falling into the Leidenfrost state. However, different from the conventional Leidenfrost state hovering by an entire air cushion (Figure 1c),[11] the drop at this state stands on the surface, held by the air cushion around the 'feet' (Figure 1b). Standing on the surface, the drop further deforms into an ellipsoid-like shape and dances like Sufi-whirling (Video S1). We captured the sequential drop images by a high-speed camera and extracted profiles. Figure 1d presents a single turn of the drop. The spherical drop evolves into an ellipsoidal shape and rotates with a period of around 23 ms for a 20 μL water drop. Here, we term $L$ and $D$ as the equatorial and polar radii of the drop, respectively. The turning angle $\theta$ stems from the sweep of the long axis from the horizontal line. We plot $\sin\theta$ and the aspect ratio $L/D$ for 5 turns in Figure 1e. The perfectly matched sinusoid indicates the drop rotates with a constant speed and the almost constant $L/D$ ratio shows the steady shape kept by the drop. The drop rotates with a constant rate of approximately 43 Hz under the demonstrational condition (Figure S3). By plotting the aspect ratio in the whole life span of the drop (Figure S4), we found the aspect ratio evolved from a close to the unit value (spherical shape) to a value close to 1.6 (ellipsoidal shape) and then decayed with the reduction of drop volume.

We first investigated the equilibrium state of this standing Leidenfrost drop. By imaging the bottom of the drop at varying temperatures, we traced the evolution of the contact region in a single experiment (Video S2). Figure 2a shows the contact length of the adhesion region as a function of time. The inset demonstrates typical images during the evolution. The change of contact length falls into five stages. When a drop is placed on this hot hydrophobic surface, nucleating boiling occurs at the touch of the surface with bubble generation until the full touch of the drop (Stage I). Because of the sudden decrease of the surface temperature by the drop at the touching area, the boiling is mild in this stage (Video S2). With the recovery of the temperature, the boiling then becomes drastic as the bubble grows and merges (Stage II). The mobility of the bubble was controlled by the wettability of the surface. For the hydrophobic surface, the growing bubble easily escaped from the side of the drop and the apparent contact line shrinks with a dramatic reduction of the contact line. With the continuous growth of temperature, the merging bubble forms a partial film underneath the droplet (stage III), occupying the major part of the bottom. The constraint of the adhesion, the increased pressure, and the surface tension yield a concave shape at the central bottom of the drop at the end of stage III. Since the vapor underneath the drop is not able to fully hold up the droplet, partial adhesion is found at the margin of the concave shape. The drop 'stood' on the surface with multiple feet around the bottom air cushion (inset, Figure S5). A further increase in the temperature of the substrate slightly changes the equilibrium of the drop. The contact line was self-adjusted to balance the air pressure and gravity (stage IV), finally resulting in a stable standing Leidenfrost droplet with partial adhesion (stage V, inset schematic).

The equilibrium state of this standing Leidenfrost drop can be described by the force balance which is affected by the evolution of the surface temperature. Figure 2b shows the balance of the gravitational force ($F_g$), the adhesion force ($F_a$), and the force that originates from the pressure of the air cushion ($F_p$) for a standing Leidenfrost drop. $F_g$ is calculated from the mass of the drop (calibration of the mass change in Figure S6). The adhesion force is calculated as

$$F_a = \int_0^L \gamma cos\theta_s dx = \gamma L cos\theta_s$$

where $\gamma$ is the surface tension, $\theta_s$ is the contact angle, and L is the perimeter of the contact area (contact length). Assuming that $\gamma = 56\ mN/m$ and $\theta_s = 110^0$ keep constant during the boiling, $F_a$ is determined by the evolution of the contact length (Figures 2a). The lifting force generated from the pressure of the air cushion is calculated as [4]

$$F_p = \Delta P S_{bottom}$$

Here, $\Delta P$ is the pressure difference between the air underneath the drop and the atmospheric pressure. $S_{bottom}$ is the projected area where the air cushion holds the drop. As the boiling process develops, the $F_p$ accumulates along with the liquid-air surface and grows to counteract the gravitational force and adhesion. Figure 2c plots the evolution of $F_a$ and $F_g$. $F_g$ slightly decreases with time, at a rate of 10 µN/s. The trend of $F_a$ is identical to the change of contact length, evolving to a constant value of around 120 µN within one second. The values of the $F_a$ and $F_g$ are in the same order of magnitude, indicating that both forces play roles in reaching the equilibrium state. As the area of the air cushion grows, the $F_p$ increases to counteract the resultant force of $F_a$ and $F_g$ in the vertical direction, finally reaching a standing Leidenfrost state. The force balance indicates that the evolution of the contact area places a dominant role in reaching this equilibrium state. Since the phase transition at the water-surface interface is dependent on the surface temperature ($T_i$), we then focus on the change of $T_i$ during the process.

When a water drop is dispersed on the surface, the $T_i$ is calculated by the analysis of the heat conservation between the drop and the substrate (inset in Figure 2d and Figure S7)

$$T_i = (T_w + T_s \cdot e_s/e_w)/(1 + e_s/e_w)$$

where $T_w$ is the initial temperature of the water drop. $T_s$ is the initial temperature of the substrate, $e_s$ and $e_w$ are the thermal effusivity of the substrate and the water, respectively. The detailed calculation can be found in Supplementary note 2. For the given temperature of surface and drop, the $T_i$ is determined by the ratio of thermal effusivity between substrate and water ($e_s/e_w$). For small $e_s/e_w$, $T_i$ is close to the water, while $T_i$ is close to the substrate for large $e_s/e_w$. Here, we have selected three materials with typical $e_s$, namely, silica, ceramic, and glass. The thermal properties of common materials are listed in Table S1. The temperature was characterized by an infrared camera, which was mounted below the surface (Figure S1). Figure 2d shows the typical $T_i$ change when a drop of 20 μL (20 ºC) was deposited on a 60 ºC substrate (IR images listed in Figure S8). For the silica surface, the $T_i$ has a shallow drop and then quickly recovers back to the initial temperature. $T_i$ on the glass surface, however, drops significantly and hardly recovers. For the ceramic surface, the $T_i$ decreases to a certain value immediately and then gradually increases to a threshold value. The same trend was also evidenced when water drop was placed on a hotter ceramics surface, where a standing Leidenfrost state was observed (Figure S9). The time scale of the temperature drop and increase matches the evolution of the contact length. The characterization of the $T_i$ enables the mild change of the contact length that allows the drop to adjust the force balance until an equilibrium state with a standing Leidenfrost drop. As shown in the phase diagram (Figure 2e), a proper adjustment of the initial temperature and $e_s/e_w$ enables a standing Leidenfrost drop with partial adhesion on the surfaces.

We then investigate the origin of the horizontal rotation after the Leidenfrost drop stands on the surface. To visualize the inner flow of the drop, we employed particle image velocimetry (PIV) measurement at the central plane of the drop (1 mm from the bottom).[4,12,13] As shown in Figure 3a and in Video S3, inner circulation occurs before the deformation of the drop into the ellipsoid shape. The angular velocity was extracted from the flow field and plotted in Figure 3b. A graduate increase in the circulation was found before the asymmetric deformation of the drop. The angular velocity reached 270 rad/s, which is identical to the rotation velocity of the deformed drop. The symmetric flow then turned to asymmetric deformation of the drop.

The rotation of a freestanding drop driven by inner circular flow has been a subject of long-standing interest since it is related to various phenomena ranging from atomic nuclear fission to planetary rotation. A number of experimental and theoretical investigations have been done to seek the origin and evolution of such asymmetric deformation, based on weightless fluids.[14-16] The circulation of the inside matter contradicting the self-attraction constraints stimulated a series of shapes once the symmetry breaks with the increase of the momentum potential. To confirm that the ellipsoid deformation originates from the inner circulation, we plot the *L/D* and the angular velocity (*ω*) during the rotation (Figure 3c). Despite the change of the *L/D*, which indicates the evolution of the drop shape, the ω almost keeps constant. A critical characteristic angular velocity ($\omega^* = \sqrt{\rho\omega^2 R^3/8\gamma}$) exists when the symmetry of the freestanding drop breaks and starts to rotate.[14] Here, $\rho$, $\gamma$ are the density and surface tension of the liquid, respectively. *R* is the radius of the droplet. We have confirmed this by inputting the parameters from the experimental conditions with different drop sizes (Figure 3d). The calculated $\omega^*$ are close to the critical value ($\omega^* \sim 0.56$) reported in the literature, despite the different sizes of droplets used in the experiment. The standing Leidenfrost drop provides a facile platform to study symmetry breaking by self-rotation.

Since the drop rotation is triggered by the inner rotating flow, another question may arise: How the horizontal rotation starts inside the drop? Inner rotation in Leidenfrost drop has been extensively investigated in previous years, focusing on vertical direction.[4,17] A heated drop levitated by the underneath air flow bears three gradients that induce inner flow: 1. Heat gradient inside the drop induces the buoyancy that drives the liquid circulation; 2. Heat-induced Marangoni flow at the surface that drives the dinner flow; 3. Pressure gradient-driven airflow places a shear on the drop that drives the surface flow like Marangoni flow. For a symmetric drop, these three forces outcome a vertical flow, either consisting of counter-rotative convective cells or a unique rotating cell, depending on the size of the droplet. By imaging the vertically central plane (Figure 3e) and the horizontal plane at *h*=0.5 mm (Figure S10), we found that a transition from vertical flow to horizontal rotation occurs. This means a horizontal gradient or shear force is applied to the drop. We attribute the horizontal circulation to partial adhesion. On the one hand, partial adhesion changes the vertical heat-gradient-driven flow from two cells to a single cell. For a gravity-flattened drop ($R>1.5$ mm), the constraint in the horizontal direction remains weaker than that in the vertical direction. Thus, the circulation turns to the horizontal plane. On the other hand, it changes the airflow direction underneath the drop, therefore deviating the airflow. A shear flow breaks the symmetric outflow from the center and the tangent flow develops to drive the horizontal circulation inside the drop. We have confirmed that the circulation at the bottom of the drop fast develops the inner rotation of the whole drop, using a numerical simulation (Figure S11). Once the angular rate reaches a critical value, the symmetry of the drop starts to break.

We further analyze how the partial adhesion deviates the air flow underneath the drop. Since the drop was held by the increased air pressure, a concave shape develops at the bottom of the drop. The adhesion parts often exist at the perimeter of the air cushion. This leads to a shear flow along the outer line of the air cushion. Such shear flow deviates the vertical circulation to the horizontal direction. Finally, horizontal circulation was observed inside the drop and stimulated the rotation of the drop once the angular velocity exceeded a critical value. We verified the airflow-induced horizontal circulation using an air-blowing device (Video S4). Using hydrophilic-hydrophobic patterns along a hole, the drop undergoes horizontal circulation under controlled airflow. It should be noted that asymmetric air-flow-driven circulation may also occur for Leidenfrost drop with an entire air cushion, if the drop was placed at a concave surface and an asymmetric shape exists at the bottom of the drop.[2,18]

Finally, we demonstrated the control of the standing Leidenfrost state and the rotation of drop. The analysis of the force and thermal balance for standing Leidenfrost drops drives us to consider the asymmetric substrate with distinct effusivity. By patterning materials with low $e_s$ on a substrate with high $e_s$, we can tailor the $T_i$ on the surface when a drop is dispersed on it (Figure 4a). Therefore, the adhesion force along the patterning region is sufficient to seize the Leidenfrost drop, with the rest of the bottom filled with air cushion. To generate asymmetric airflow, engineered patterns mimicking the self-adjusted contact area are adopted (Figure 4b). We also designed artificial patterns with a similar size to the contact area and in asymmetric shape (Figure 4b). When the substrate is heated over Leidenfrost temperature (e.g., 250 ºC), a stable standing Leidenfrost state was observed and even multiple rotating drops can be achieved (Figure 4c).

Experimental

The hydrophobic substrate samples were placed on a heater with a hole (3cm in diameter) for visualization at the bottom. The surface renders an apparent contact angle of 110º. A high-speed camera (Photron) was set to capture the status of the drop from the top, bottom, and side, respectively. For temperature measurement, an IR camera (Fortic) was used to visualize the surface temperature distribution from the bottom. The filming speed ranges from 500 fps to 2000 fps, depending on the motion of the drop. Internal flows are visualized using a home-built PIV system. Tracers (Polystyrene, 4 μm in diameter) were dispersed in water. To illuminate the particles, a light sheet with a thickness of 50 μm was generated from a laser (wavelength =532). Images captured at a speed of 2000 fps were further analyzed by PIVlab to extract the velocity field inside the drop.

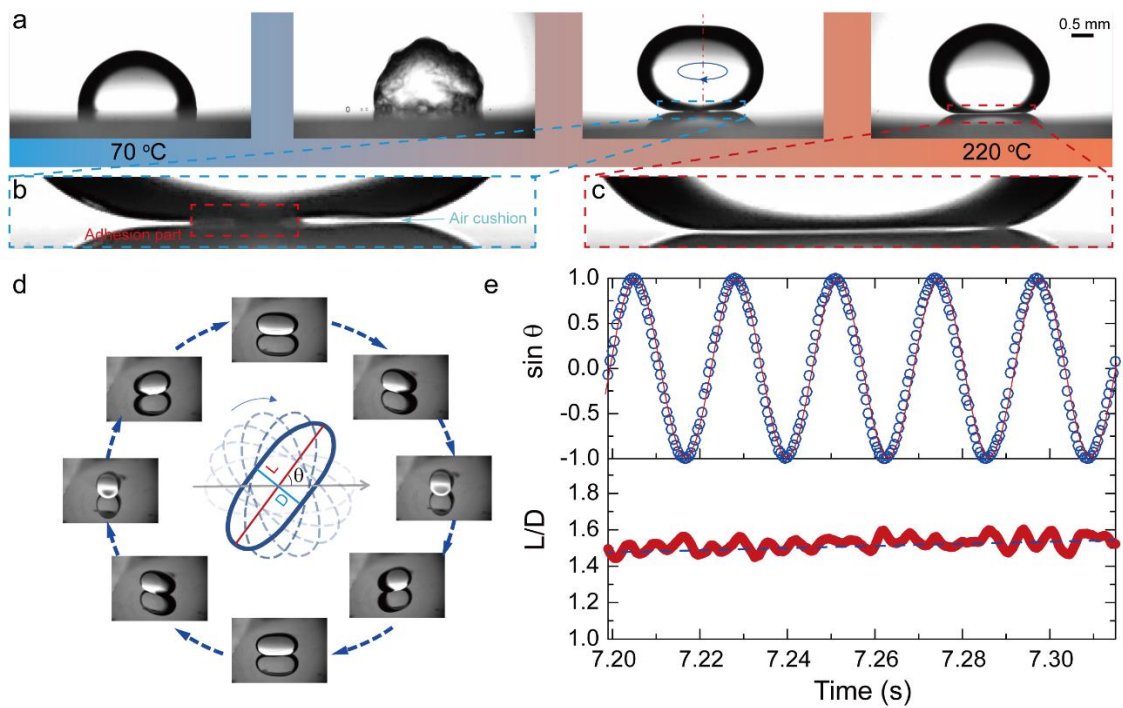

**Figure 1** Sufi-whirling Leidenfrost drops. (a) Water drops (20 μL) were gently deposited on a heated hydrophobic surface at varying temperatures. The drop either undergoes dramatic boiling or keeps stable hovering by an air cushion (b and c). In addition to conventional boiling phenomena, a stable Leidenfrost state was found with partial adhesion (b) to the surface. The drop may evolve to horizontal deformation and rotate like Sufi whirling at this state. (d) A single rotation period of a rotating Leidenfrost drop. The drop deforms to an elliptical shape and rotates with a typically period of 23 ms for the volume of 20 μL. (e) Variation of the turning angle $\theta$ and the aspect ratio $L/D$ with time in 5 periods for a rotating drop with volume of 20 μL. $\theta$, $L$ and $D$ are defined in (d). The perfectly matched sinusoid indicated the drop rotates with a constant speed and the almost constant $L/D$ ratio shows the steady shape kept by the drop.

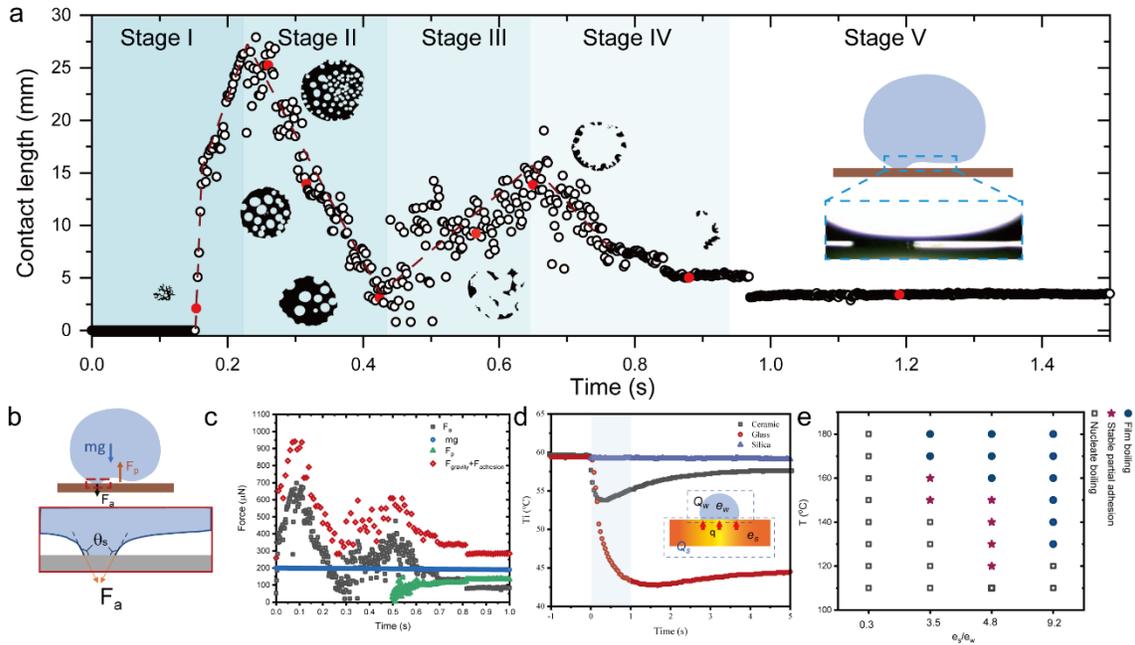

**Figure 2** Origin of the stable Leidenfrost state with partial adhesion. (a) Evolution of the contact length for a stable Leidenfrost drop with partial adhesion. The change of contact length typically falls into five stages. The drop undergoes nucleating boiling and transitions to stable filming boiling with stable partial adhesion like a standing Leidenfrost state. The inset shows the stable Leidenfrost drop with air cushion underneath and partial adhesion. (b) Force balance for a stable Leidenfrost drop with partial adhesion. The levitating force by the air pressure balances the gravitational force and adhesion. (c) Evolution of the pressure force, the gravitational force and the adhesion force. (d) Typical surface temperature change when a water drop (20 $^{o}$C) placed on the surface for materials with high (silica), moderate (ceramic), and low (glass) thermal effusivity. (e) The effect of temperature and the ratio of thermal effusivity on the drop boiling. Leidenfrost state with stable partial adhesion was only found for moderate ratio of thermal effusivity.

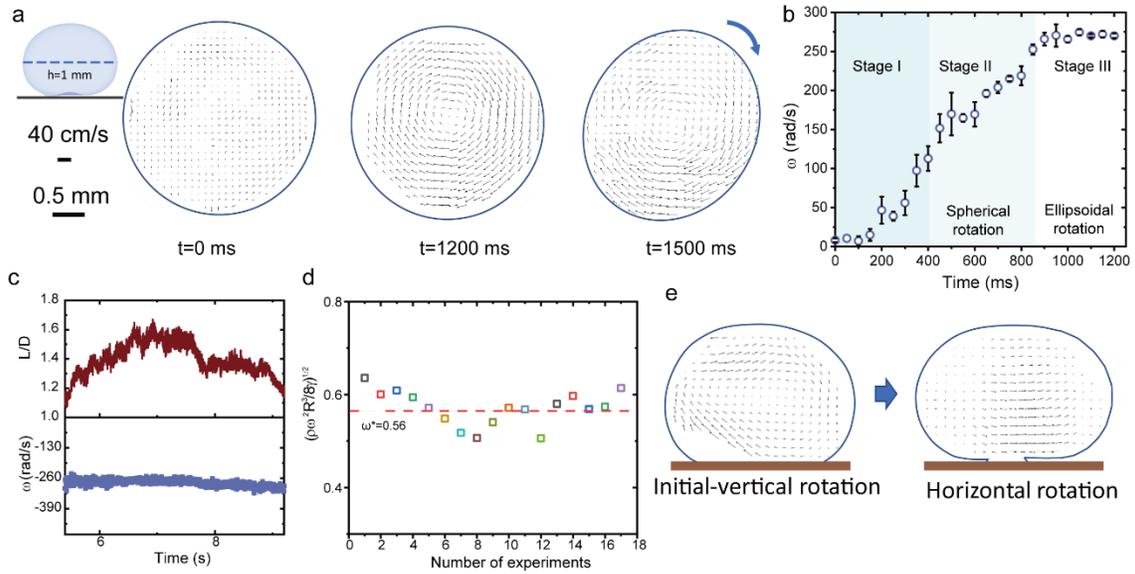

**Figure 3** Origin of the deformation and rotation of the Leidenfrost drop with stable partial adhesion. (a) PIV measurement shows the evolution of the inner flow at the horizontal central plane (1 mm from the bottom). Tracers reveal circulation first occurs inside the drop before the deformation of the drop. (b) Rotating velocity (ω) as a function of time. ω increases from zero (initial) to a critical value when drop deforms to ellipsoid. At the increasing period, the drop keeps spherical despite the inner rotation. (c) The constant rotating velocity confirms the inner rotation-induced deformation. (d) Characteristic angular velocity. (e) PIV measurement shows the evolution of inner flow at the vertical central plane. Vertical circulation finally evolves to horizontal circulation for a standing Leidenfrost drop with partial adhesion.

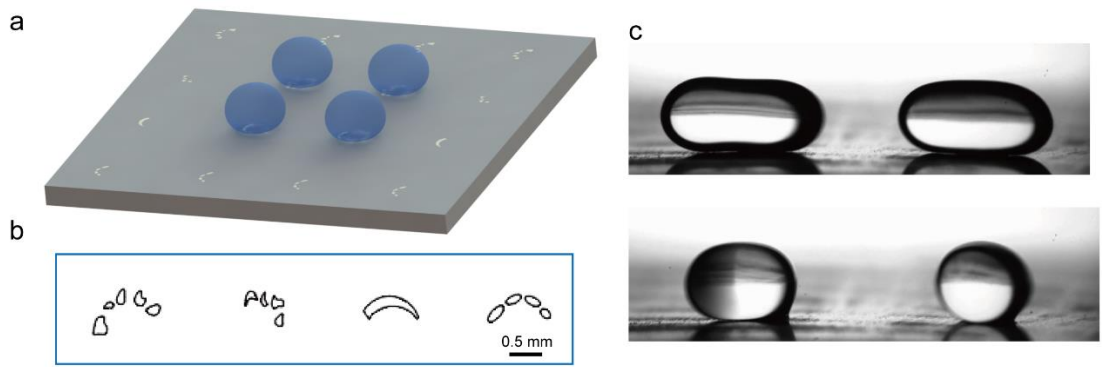

**Figure 4** Engineered patterns for the manipulation of standing Leidenfrost drops and rotation. (a) Engineered surface introduces anisotropic $e_s$ at the surface. By specifically designing the shape of the pattern (b), asymmetric air flow is achieved at the bottom of the drop and drives the drop to rotate. The demonstrated patterns originate from the partial adhesion on smooth hydrophobic surface (left two in b) and artificial asymmetric shape (right two in b). (c) Multiple standing Leidenfrost drop and rotation can be achieved by the anisotropic patterns.